\documentclass[aps,twocolumn,eqsecnum,bibnotes,showpacs]{revtex4}
\usepackage{graphicx}
\usepackage{bm}
\usepackage[tbtags]{amsmath}

\def\half{\mbox{\small $\frac{1}{2}$}}

\def\beq{\begin{equation}}
\def\eeq{\end{equation}}
\def\eeql#1{\label{#1} \end{equation}}
\def\f{\bm{f}}

\def\ot{\! \otimes \!}
\def\om{\omega}
\def\H{{\cal H}}
\def\bdot{{\mbox{\boldmath $\cdot$}}}
\def\im{\mathop{\rm Im}\nolimits}
\def\re{\mathop{\rm Re}\nolimits}
\def\diag{\mathop{\rm diag}\nolimits}
\newcommand{\phs}{{\vphantom{*}}}

\begin{document}


\title{Hamiltonian and Linear-Space Structure for Damped Oscillators: I. General Theory}

\author{S.C. Chee}
\author{Alec \surname{Maassen van den Brink}}
\thanks{Corresponding author;\\ electronic address: \texttt{alec@dwavesys.com}}
\author{K.~Young}
\affiliation{Physics Department, The Chinese University of Hong Kong, Hong Kong, China}

\date{first posted on 18 Jun 2002; revised on 10 Feb 2004}


\begin{abstract}
The phase space of $N$ damped linear oscillators is endowed with a bilinear map under which
the evolution operator is symmetric. This analog of self-adjointness allows properties familiar from conservative systems to be recovered, e.g., eigenvectors are ``orthogonal" under the bilinear map and obey sum rules, initial-value problems are readily solved and perturbation theory applies to the \emph{complex} eigenvalues. These concepts are conveniently represented in a biorthogonal basis.
\end{abstract}


\maketitle


\section{Introduction}
\label{sect:intro}

Conservative systems are described by self-adjoint time-evolution operators $\H$ on a linear space, with well-developed mathematical tools. Eigenvectors are orthogonal and complete, and eigenvector expansions are unique, with coefficients given by projection. Expansion of the initial data solves the dynamics, with each term simply acquiring a phase $e^{-i\om_j t}$, where $\om_j$ are the eigenvalues.

Indeed much of our very language, e.g., normal modes, energy levels, transition frequencies, photons and phonons, is rooted in eigenvector expansions.  These examples already suggest that the present ostensibly classical problems are also relevant in the quantum domain.

However, these concepts are ruined by dissipation, since $\H$ is no longer self-adjoint.  This paper shows, for ohmically damped oscillators, that many properties well known from the conservative case can be resurrected, in terms of not the standard inner product, but a bilinear map under which $\H$ turns out to be symmetric.  This symmetry is analogous to self-adjointness, and leads to the orthogonality of eigenvectors under the bilinear map, the usual ready solution of initial-value problems and Rayleigh--Schr\"odinger perturbation theory (RSPT) in the familiar form but applicable to complex eigenvalues.

Serious microscopic discussion of dissipation began with a single linear oscillator coupled to an infinite bath, which is then eliminated from the equations of motion or path integral~\cite{ref1}. A huge body of works generalize this to many nonlinear coupled oscillators~\cite{ref2}. Here, however, we first deal with $N$ coupled \emph{linear}, \emph{classical} oscillators, with coordinates $\phi(\alpha)$, $\alpha = 1, \ldots, N$, described by
\beq
  d_t^2 \phi(\alpha) + \Gamma(\alpha , \beta) d_t \phi(\beta) 
  + K(\alpha , \beta) \phi(\beta) = 0
\eeql{eq:eqmot1}
(summation convention for Greek indices), with $\Gamma$ a symmetric non-negative damping matrix and $K$ a symmetric positive matrix of force constants~\cite{mass}. One can obtain (\ref{eq:eqmot1}) by elimination of a bath, giving the dynamics the necessary quantum pedigree; but until Section~\ref{sect:disc}, it can simply be regarded as phenomenological. The non-negativity of $\Gamma$ may be relaxed, e.g., in continuum models of the electromagnetic field in \emph{gain} media.

Section~\ref{sect:form} sets up the linear-space formalism. However, eigenvector expansions would not be useful unless there is a convenient projection; for this purpose, the key bilinear map $(\bm{\psi},\bm{\phi})$ between state vectors in $2N$-dimensional phase space (leading to a biorthogonal basis) is introduced; examples are given in Section~\ref{sect:ex}. The application to time-independent perturbation theory (Section~\ref{sect:pert}) for the \emph{complex} eigenvalues and eigenvectors proceeds by mimicking RSPT for conservative systems. Section~\ref{sect:cont} briefly discusses the continuum limit $N\rightarrow\nobreak\infty$, recovering results for waves in open systems, which have been studied extensively~\cite{openwavermp}.

Section~\ref{sect:disc} sketches the application to thermal and quantum physics, and to nonlinear oscillators. As a further outlook, Appendix~\ref{sect:constr} considers the separation of time scales if one oscillator is light. Adiabatic elimination of its momentum yields effective constrained dynamics, giving an example where the methods still apply also when phase space is \emph{odd}-dimensional. An extensive account can be found in Ref.~\cite{chee}.

The formalism assumes that (a)~although $\H$ is not self-adjoint, its eigenvectors $\f_j$ are complete, and (b)~for all~$j$, $(\f_j,\f_j)\neq0$. These are in fact related, and violated only at \emph{critical points} (with measure zero in parameter space) where eigenvectors merge (Section~\ref{subsect:crit} and Ref.~\cite{pap2}). A brief treatment focusing on ``excess noise'' in open systems can be found in Ref.~\cite{Pet}.


\section{Formalism}
\label{sect:form}

\subsection{State space and evolution operator}
\label{subsect:stsp}

We cast (\ref{eq:eqmot1}) in first-order form by introducing the momenta $\hat{\phi}(\alpha)$. Denote the state in phase space as
\beq
  \bm{\phi} = (\phi,\hat{\phi})^{\rm T}
  = (\phi^1, \ldots, \phi^{2N} )^{\rm T} = \sum_{j=1}^{2N} \phi^j \bm{u}_j\;,
\eeql{eq:state}
where $\bm{u}_j$ are the basis vectors of this canonical representation, and boldface denotes $2N$-vectors. Then
\begin{gather}
  d_t \bm{\phi} = -i \H \bm{\phi}\;,\label{eq:eqmot3}\\
  \H^\bdot{}_\bdot = i \begin{pmatrix} 0 & I \\ -K & -\Gamma \end{pmatrix}\,,\label{eq:defh}
\end{gather}
with $I$ the $N \times N$ identity. The dots specify whether the indices are contravariant or covariant (cf.\ Section~\ref{subsect:matrep}).

Eigenvalues are the roots of
\beq
  J(\om) \equiv \det(\H^\bdot{}_\bdot - \om)\;,
\eeql{eq:charpoly}
and right eigenvectors $\f_j$ are defined by
\beq
  \H \f_j = \om_j  \f_j \;.
\eeql{eq:eigenright}
Left eigenvectors $\f^j$ will be introduced later.  For $\f_j$, (\ref{eq:eqmot3}) implies the time-dependence $e^{-i\om_j t}$, so
\beq
  \hat{f}_j= -i\om_j f_j\;,
\eeql{eq:compon2}
and one can simply solve (\ref{eq:eqmot1}) in the frequency domain for the coordinates, without reference to the momenta:
\beq
  [-\om_j^2 \delta(\alpha , \beta) -i\om_j \Gamma(\alpha , \beta)
  + K(\alpha , \beta) ] \, f_j(\beta) = 0\;.
\eeql{eq:eqmot4}

With dissipation, $\im\om_j \le 0$~\cite{damp}. Conjugating (\ref{eq:eigenright}), it follows that $\om_{-j}^\phs=-\om_j^*$ is also an eigenvalue, with $\f_{-j}^\phs \propto \f_j^* $ (see below for normalization and phase). Thus, eigenvalues are paired, except for so-called zero-modes with $\re\om_j=0$~\cite{zero-pair}. Indeed, the determinant of $[\cdots]$ in (\ref{eq:eqmot4}) has $2N$ roots $\om_j$; if $\Gamma\neq0$ then typically also all $\om_j^2$ are different, showing that an eigenvector expansion can only be sought in phase, not coordinate, space.

\subsection{Inner product and bilinear map}
\label{subsect:map}

In phase space, the standard inner product is~\cite{timescale}
\beq
  \langle \bm{\psi} | \bm{\phi} \rangle =  \psi(\alpha)^* \phi(\alpha) 
  + {\hat \psi}(\alpha)^* {\hat \phi}(\alpha)\;,
\eeql{eq:inner}
with $\langle \bm{\phi} | \bm{\phi} \rangle \ge 0$. Unfortunately, 
$
\langle \bm{\psi} | \H \bm{\phi} \rangle \neq
\langle \bm{\phi} | \H \bm{\psi} \rangle^*
$, so $\langle \f_k | \f_j \rangle\neq 0$ even if $\om_k \neq \om_j$.

Our key concept is the bilinear map:
\begin{align}
  ( \bm{\psi} , \bm{\phi} )&\equiv i \bigl[ \psi(\alpha) {\hat \phi}(\alpha)
    + {\hat \psi}(\alpha) \phi(\alpha) + 
    \psi(\alpha) \Gamma(\alpha , \beta) \phi(\beta) \bigr]\notag \\
  &=( \bm{\phi} , \bm{\psi} )\;.\label{eq:blmap}
\end{align}
The diagonal entries $(\bm{\phi},\bm{\phi})$ are not positive definite, not even necessarily real~\cite{real}. By (\ref{eq:compon2}), $(\f_j,\f_k)$ can be expressed in terms of the coordinates alone.

We will justify (\ref{eq:blmap}) by the ensuing properties, crucially (\ref{eq:symh}) below; still, let us motivate it immediately. First consider the conservative case $\Gamma=0$.  A structure $\psi\phi + \hat{\psi}\hat{\phi}$ is not dimensionally correct~\cite{timescale}, but $\psi\hat{\phi}  + \hat{\psi}\phi$ is.  Moreover, with (\ref{eq:blmap}) an eigenvector pair with coordinates $f_j(\alpha,t)=f_j(\alpha) e^{-i\om_jt}$ and $f_{-j}(\alpha,t)=f_j(\alpha)^* e^{i\om_jt}$ will have opposite momenta, hence a zero bilinear map. This  would not hold if one vector is conjugated. When there is damping, the last term $\psi \Gamma \phi$ is the only other type with the correct dimension; with a unit coefficient as in (\ref{eq:blmap}), it ensures
\beq
  ( \bm{\psi} , \H \bm{\phi} ) = ( \bm{\phi} , \H \bm{\psi} )\;.
\eeql{eq:symh}
In the proof, $\Gamma$ in $\H$ cancels $\Gamma$ in~(\ref{eq:blmap}). This analog of self-adjointness, as usual, yields ``orthogonality"
\beq
  ( \f_j , \f_k ) = 0\qquad\mbox{if $\om_j \neq \om_k $}\;,
\eeql{eq:orth}
including pairs $\om_{-j}$, $\om_j$. Taking suitable linear combinations, (\ref{eq:orth}) can be extended to level crossings, i.e., multiple independent eigenvectors at one eigenvalue.

\subsection{Normalization and phase}
\label{subsect:norm}

If $(\f_j , \f_j) \ne 0$ (cf.~Section~\ref{subsect:crit}), completeness implies
\beq
  \bm{\phi} = \sum_j \frac{(\f_j,\bm{\phi})}{(\f_j, \f_j)}\,\f_j\;.
\eeql{eq:comp0a}
This normalization-independent form is appropriate if small denominators must be handled for $(\f_j, \f_j) \approx\nobreak 0$. Also, the usual convention for the conservative limit is $f_j(\alpha) f_j(\alpha) = \nobreak1$, i.e., $(\f_j , \f_j) = 2\om_j$ [incidentally explaining the conventional factor~$i$ in (\ref{eq:blmap})]. Thus, connection with this limit is best expressed via~(\ref{eq:comp0a}).

However, in most other circumstances, it is more convenient to adopt the normalization (and phase) convention $( \f_j , \f_j) = 1$ for all $j$, in which case
\beq
  (\f_j , \f_k) = \delta_{jk}\;.
\eeql{eq:orth1}
Unless otherwise specified, expressions that are not manifestly normalization-independent conform to (\ref{eq:orth1}).

Conjugate eigenvectors have the phase relationship
\beq
  \f_{-j}=\pm i \f_j^*
\eeql{eq:norm3}
in order to satisfy (\ref{eq:orth1})~\cite{deg-phase}.

\subsection{Metric}
\label{subsect:matrep}

In the $\bm{u}$-basis~\cite{notation}, $(\bm{\psi},\bm{\phi})=\sum_{i,j}\psi^ig_{ij}\phi^j$ in terms of 
\beq
  g_{ij} = ( \bm{u}_i , \bm{u}_j )\;,\qquad
  g_{\bdot\bdot} = i \begin{pmatrix} \Gamma & I \\ I & 0 \end{pmatrix}\,.
\eeql{eq:defg3}
This, and the inverse metric 
\beq
  g^{\bdot\bdot} = -i \begin{pmatrix} 0 & I \\ I & -\Gamma \end{pmatrix} 
\eeql{eq:defg4}
are used to lower/raise indices as usual. This explicit form shows that $g_{\bdot\bdot}$ is non-singular. Moreover, if $(\bm{\psi} , \bm{\phi} ) = 0$ for every $\bm{\psi}$, then $0 = ( g^{\bdot\bdot} \bm{\phi}^* , \bm{\phi} ) = \sum_j |\phi^j|^2 $, implying $\bm{\phi} = 0$. In the eigenbasis, $\bar{g}_{ij} = ( \f_i,\f_j ) = \delta_{ij}$.

The crucial symmetry (\ref{eq:symh}) translates into the symmetry of $\H_{\bdot\bdot}=g_{\bdot\bdot}\,\H^\bdot{}_\bdot$\;. Explicitly,
\beq
  \H_{\bdot\bdot} = \begin{pmatrix} K & 0 \\ 0 & -I \end{pmatrix}\,,
\eeql{eq:defh3}
in the evaluation of which [cf.~(\ref{eq:defh}) and (\ref{eq:defg3})] the cancellation of $\Gamma$ becomes apparent. Incidentally, $\H_{\bdot\bdot}$ can be expressed directly in terms of the matrix elements $(\bm{\psi},\H\bm{\phi}) = \sum_{i,j} \psi^i \H_{ij} \phi^j$, and using (\ref{eq:defh3}) gives
\begin{align}
  ( \bm{\phi},\H \bm{\phi} )&= -{\hat \phi}(\alpha){\hat \phi}(\alpha)
  + \phi(\alpha) K(\alpha,\beta) \phi(\beta)\notag \\
  &= -2L[\bm{\phi}]\;,\label{eq:lag}
\end{align}
where $L$ can be interpreted (at least in the non-dissipative limit) as the Lagrangian.

Although $\Gamma$ disappears from (\ref{eq:defh3}), physical quantities of course do depend on the damping.  For example, $\om_j$ is the stationary value of $(\bm{\phi},\H\bm{\phi})/(\bm{\phi},\bm{\phi})$ attained when $\bm{\phi}=\f_j$.  The numerator (\ref{eq:lag}) is $\Gamma$-independent; but the denominator carries the nontrivial dependence on $\Gamma$. 

\subsection{Duality}
\label{subsect:dual}

For any basis $ \{\bm{v}_j\} $, there is a dual $\{\bm{v}^j\}$ such that $\langle \bm{v}^j | \bm{v}_k \rangle = \delta^j{}_k$~\cite{biorth}. For the canonical basis, $\bm{u}^n = \bm{u}_n$, so that the $2N \times 2N$ coefficient matrices (in the $\bm{u}$-basis) $(v_j)^n$ and $(v^j)^*_n$ are each other's inverse. However, for the eigenbasis one can bypass this inversion, since (\ref{eq:orth1}) and (\ref{eq:defg3}) give $(f^j)_m = \sum_n[ g_{mn} (f_j)^n]^*$, that is,
\beq
  \f^j = {\cal D} \f_j \equiv {[ g_{\bdot\bdot} \f_j ]}^*\;.
\eeql{eq:dualmap}
This is extended to any vector $\bm{\phi}$, i.e., ${\cal D} \bm{\phi} \equiv [ g_{\bdot\bdot}\, \bm{\phi}]^*$.

The duals are left eigenvectors $\H^{\dagger} \f^j = \om_j^* \f^j$. In terms of these, $\bm{\phi}= \sum_j \f_j\, \langle \f^j | \bm{\phi} \rangle$ and, in an obvious shorthand,
\beq
  {\cal I} = \sum_j \f_j \,\langle \f^j | \, \bdot \, \rangle\;,
\eeql{eq:comp2}
with ${\cal I}$ the $2N \times 2N$ identity. The dual basis leads to a consistent use of contravariant and covariant indices.

\subsection{Sum rules}
\label{subsect:sum}

Writing out (\ref{eq:comp2}) in components yields four sum rules:
\begin{subequations}
\begin{align}
  0 &= \sum_j f_j \ot f_j \;, \label{eq:sr1} \displaybreak[0]\\
  I &= \sum_j \om_j f_j \ot f_j \;, \label{eq:sr2} \displaybreak[0]\\
  0 &= \sum_j f_j \ot \, (\Gamma f_j)\;, \label{eq:sr3} \displaybreak[0]\\
  0 &= \sum_j [\om_j^2 f_j \ot  f_j + i \om_j f_j \ot \, (\Gamma f_j) ]
  \;. \label{eq:sr4}
\end{align}
\end{subequations}%
Here, $a \ot b$ stands for the matrix with elements $a(\alpha) b(\beta)$. We have verified all these in examples.

For $\Gamma \rightarrow 0$, (\ref{eq:sr1}), (\ref{eq:sr3}) and (\ref{eq:sr4}) become vacuous. Namely, all $\re\om_j \neq 0$ for small $\Gamma$, so the $f_j$ are paired, obeying (\ref{eq:norm3}). Thus, e.g., $f_{-j}(\alpha) f_{-j} (\beta) \rightarrow - f_j(\alpha) f_j(\beta)$, making (\ref{eq:sr1}) trivial. The remaining (\ref{eq:sr2}) reduces to the familiar sum rule for conservative systems (cf.~Section~\ref{subsect:norm} for the different normalizations).

\subsection{Time evolution}
\label{sect:time}

In terms of (\ref{eq:comp0a}), the dynamics is formally solved as
\beq
  \bm{\phi}(t) = \sum_j\phi^j e^{-i\om_j t} \f_j\;,\qquad
  \phi^j = \bm{(}\f_j,\bm{\phi}(t{=}0)\bm{)}\;.
\eeql{eq:expandt}
This familiar construction would \emph{not} be possible with the product (\ref{eq:inner}), which would have required inverting the $2N \times 2N$ matrix $\langle \f_j | \f_k \rangle$---nontrivial if $N$ is large. Of course, to use (\ref{eq:expandt}), one must know the $\f_j$, but when only a few $\f_j$'s dominate the dynamics, the burden of finding those (e.g., iteratively) is comparatively small.

In operator form [cf.\ (\ref{eq:comp2})], the Green's function is $\bm{\phi}(t) \theta(t) = {\cal G}(t) \bm{\phi}(t{=}0)$ ($\theta$ is the unit step function), or
\beq
  ( \H - \om ) \, {\tilde {\cal G}} (\om) = -i {\cal I}\;.
\eeql{eq:green02a}
With the eigenvectors complete, explicitly one has
\beq
  {\tilde {\cal G}} (\om) = \begin{pmatrix} \tilde{G}^{QQ} & \tilde{G}^{QP} \\
                            \tilde{G}^{PQ} & \tilde{G}^{PP} \end{pmatrix}
   = \sum_j \f_j \frac{i}{\om - \om_j}\,( \f_j , \bm{\bdot} )\;,
\eeql{eq:green03}
Fourier inversion of which reproduces (\ref{eq:expandt}).

Using (\ref{eq:defh}), in the canonical basis we can also write
\beq
  \tilde{\cal G}(\om) = (-\om^2 -i\om \Gamma + K )^{-1}
  \begin{pmatrix} \Gamma -i\om  & I \\ -K & -i\om \end{pmatrix}\,,
\eeql{eq:gg04}
involving the inversion of only an $N \times N$ matrix, in contrast to 
(\ref{eq:green02a}). The first factor in (\ref{eq:gg04}) corresponds to the differential operator in (\ref{eq:eqmot4}).

\subsection{Exceptions}
\label{subsect:crit}

Our formalism relies on (a)~the completeness of eigenvectors, and (b)~$(\f_j , \f_j) \neq 0$ for all $j$. These assumptions are related. Imagine that system parameters are tuned, and the $2N$ eigenvectors start off (e.g., for $\Gamma=0$) complete.  Then (a) is violated when two (or more) eigenvectors merge, e.g.\ $\f_k \rightarrow \f_j$, so $(\f_j , \f_j) = \lim \, (\f_j , \f_k) = 0$. Conversely, if (b) is violated, $\f_j$ is orthogonal to every eigenvector; if these would be complete, then $\f_j$ is orthogonal to all~$\bm{\psi}$, contradicting $\f_j \neq 0$, cf.~below (\ref{eq:defg4})~\cite{zeronorm}.

These conditions are violated only for a zero-measure set of parameters. Section~\ref{sect:ex} contains examples; the extension to these critical points will be given separately~\cite{pap2}.


\section{Examples}
\label{sect:ex}

\subsection{Single oscillator}
\label{subsect:exa}

Although the theory is most valuable when $N \gg 1$ (otherwise brute force suffices), it is already nontrivial for $N=\nobreak1$. Let $K = k$ and $\Gamma = \nobreak2\gamma$. The eigenvalue equation $J(\om)=0$ [cf.\ (\ref{eq:charpoly})] leads to $\om_{\pm} = \pm  \Omega - i\gamma$, where $\Omega = \sqrt{k - \gamma^2}$, and
\beq
  \f_{\pm} = c_{\pm} \begin{pmatrix} 1 \cr -i\om_{\pm} \end{pmatrix}\,.
\eeql{eq:exa03}
The bilinear maps are $(\f_+,\f_-)=0$ and
\beq
  ( \f_{\pm} , \f_{\pm} ) = \pm 2 c_{\pm}^2 \Omega\;.
\eeql{eq:exa05}

At the critical point $k = k_* \equiv\gamma^2$, (a) the eigenvalues merge: $\om_{+} = \om_{-} = -i \gamma$; (b) from (\ref{eq:exa03}) the eigenvectors merge, leaving only one in the 2-d space; and (c) from (\ref{eq:exa05}) the diagonal bilinear map vanishes. This example also explains why eigenvector merging is called criticality.

\subsection{Two oscillators}
\label{subsect:exb}

Let $N=2$ with~\cite{mass}
\beq
  K = \begin{pmatrix} \phantom{-}4 & -2 \\ -2 & \phantom{-}4 \end{pmatrix}\,, \qquad
  \Gamma = \begin{pmatrix} 4\gamma & 0 \\ 0 & 2\gamma \end{pmatrix}\,.
\eeql{eq:exb01}
As $\gamma$ increases, one mode pair becomes overdamped for $\gamma > \gamma_{*1} = 0.8599$, and the other for $\gamma > \gamma_{*2} =2.1031$.  Except at these critical points, the eigenvectors are complete, and all bilinear maps etc.\ can be found explicitly. A perturbation of (\ref{eq:exb01}) will be studied in Section~\ref{subsect:ptnd}.

\subsection{Chain of masses}
\label{subsect:exc}

Next consider a chain of unit masses with displacements $\phi(\alpha)$ from the rest positions $x_\alpha = \alpha \Delta$. Mass $\alpha$ is tied to $x_\alpha$ and its nearest neighbors by springs of force constant $V(\alpha)$ and $k$ respectively, and is moreover subject to a damping force $-\Gamma(\alpha) \partial_t \phi(\alpha)$. That is,
\begin{align}
  K(\alpha, \beta) &= \left\{\!\! \begin{array}{ll}
  2k + V(\alpha) & \mbox{if $|\alpha - \beta| = 0$} \\
  -k & \mbox{if $|\alpha - \beta| =1$} \\
  0 & \mbox{if $|\alpha - \beta| >1$}
  \end{array} \right.\;,\notag \\
  \Gamma(\alpha,\beta) &= \Gamma(\alpha) \delta(\alpha, \beta)\;.\label{eq:stringa01}
\end{align}
In the obvious continuum limit $\Delta \rightarrow 0$, $k\Delta^2 = 1$, the system obeys the (generalized) Klein--Gordon equation
\beq
  [ \partial_t^2  + \Gamma(x) \partial_t 
  - \partial_x^2 + V(x)] \phi(x,t) = 0\;,
\eeql{eq:stringa02}
discussed in Section~\ref{sect:visc}.

Let $0\le\alpha\le N{+}1$, $\phi(0)=\phi(N{+}1)=0$. In the special solvable case $\Gamma(\alpha)=\nobreak 2\gamma$, (\ref{eq:eqmot4}) reads $K(\alpha,\beta) f_j(\beta)=(\om_j^2 + 2i\gamma \om_j) f_j(\alpha)$. But $K$ has a complete real eigensystem $\{\Omega_j^2,f_j\}$, with $\Omega_j$ the frequencies for $\gamma=0$. Now
\beq
  \om_j = -i\gamma \pm\sqrt{ \smash{\Omega_j^2}{-}\smash{\gamma}^2}\;,
\eeql{eq:stringa05}
while the eigenvectors remain the same. Thus this model is in effect a superposition of many oscillators, each as in Section~\ref{subsect:exa}. For any $\gamma$, at most one of these can be critical.  This example is simple because $[K,\Gamma] = 0$.

Let us comment briefly on the doubling of modes~\cite{twocomp}. For $\Gamma = 0$, the $\Omega_j^2$ are real; but each of these splits into \emph{two} complex $\om_j^2$ when $\Gamma \ne 0$. Modes are doubled because expansions now refer to $\phi$ and $\hat{\phi}$ simultaneously.

The $ \f_j $-basis for the [\emph{un}]damped system is orthogonal under (\ref{eq:blmap}) [(\ref{eq:inner})]. This is consistent, since
\beq
  (\f_j , \f_j) = \pm 2\sqrt{ \smash{\Omega_j^2}{-}\smash{\gamma}^2} \, f_j(\alpha) f_j(\alpha)
\eeql{eq:stringa06}
equals the conventional norm for the conservative system except for the prefactor, vanishing at critical damping.  

To be even more specific, let $V(\alpha)=0$, so that the undamped case is essentially the familiar model for acoustical phonons, with $\Omega_j^2 = 2k\{1 -\cos[j\pi/(N{+}1)]\}$ and $f_j(\alpha)=c_j\sin[j\pi\alpha/(N{+}1)]$ for $1\le j\le N$. Setting (\ref{eq:stringa06}) to unity, one finds $c_j=[(\om_j{+}i\gamma)(N{+}1)]^{-1/2}$.

\subsection{Inhomogeneous damping}
\label{subsect:exd}

A non-trivial solvable model with $[ K, \Gamma ] \neq 0$ can be constructed as follows. Let the masses be $\alpha = 0, \ldots , N$, with again $\phi(0)=0$ but with $\phi(N)$ free, so there are $N$ dynamical variables, with force constants as in (\ref{eq:stringa01}) except that
\beq
  K(N,N) = k + V(N)\;.
\eeql{eq:stringb00}
Let only the final mass be damped:
\beq
  \Gamma(\alpha,\beta) = \gamma\delta(\alpha,N)\delta(\beta,N)\;.
\eeql{enddamp}

We only consider $V=0$, returning to finite $V$ in Section~\ref{sect:open}. For $N\rightarrow\infty$, it is then straightforward to find
\beq
  \frac{N\om_j}{\sqrt{k}}=\left\{\!\!\begin{array}{ll}
    (j{+}\half)\pi-i\mathop{\rm artanh}(\gamma/\sqrt{k})\;, & \gamma < \sqrt{k}\;;\\[1mm]
  j \pi -i \mathop{\rm arcoth} (\gamma / \sqrt{k})\;, & \gamma > \sqrt{k}\;.
  \end{array}\right.
\eeql{eq:stringb06}
The divergence as $\gamma\rightarrow\sqrt{k}$ occurs only for $N\rightarrow\infty$. For finite $N\gg1$, (\ref{eq:stringb06}) fails very near $\gamma=\nobreak\sqrt{k}$, and the true $\om_j$ are continuous in $\gamma$. In particular, for $\gamma=\sqrt{k}$ one has $\om_j=(\sqrt{k}/N)[(j{+}\chi)\pi-\frac{1}{2}i\ln N]$, where $\chi=\frac{3}{4}$, $\frac{1}{2}$, and~$\frac{1}{4}$ for $-N\ll j\ll-\ln N$, $|j|\ll\ln N$, and $\ln N\ll j\ll\nobreak N$ respectively.


\section{Perturbation Theory}
\label{sect:pert}

The familiar RSPT can be transcribed, everywhere replacing the usual inner product with the bilinear map (\ref{eq:blmap}). However, this assumes that the bilinear map itself is unperturbed, so we only consider, for small real $\epsilon$,
\beq
  \H = \H_0 + \epsilon \Delta \H\;,\qquad
  \Delta \H = i \begin{pmatrix} 0 & 0 \\ -\Delta K & 0\end{pmatrix}\;.
\eeql{eq:ndp01}

\subsection{Simple modes}
\label{subsect:ptnd}

The changes in the eigenvectors up to first order, and in the eigenvalues up to second order, are~\cite{alternate}
\begin{align}
  \Delta \bm{f}_{j} &= \epsilon \, \sum_{k \neq j}
  \bm{f}_{k} \, \frac{( \bm{f}_{k} , \Delta \H \bm{f}_{j} )}{\om_j - \om_k}\;,
  \label{eq:ndp03}\\
  \Delta \omega_{j} &= \epsilon \, ( \bm{f}_{j} , \Delta \H \bm{f}_{j} ) +
  \epsilon^2 \, \sum_{k \neq j}
  \frac{( \bm{f}_{k} , \Delta \H \bm{f}_{j} )^{2}}{\om_j - \om_k}\;.\label{eq:ndp04}
\end{align}
In the $\epsilon^2$-term, the matrix elements are squared \emph{without} taking absolute values.  Higher orders are easily written down. Here and below, $\{\om_j,\f_j\}$ is the unperturbed eigensystem. In view of (\ref{eq:ndp01}), one simply has
\beq
  ( \f_k , \Delta \H \f_j ) = f_k(\alpha) \Delta K(\alpha,\beta) f_j(\beta)
  \equiv (\Delta K)_{kj}\;.
\eeql{eq:ndp05}

\begin{figure}
  \includegraphics[width=8cm]{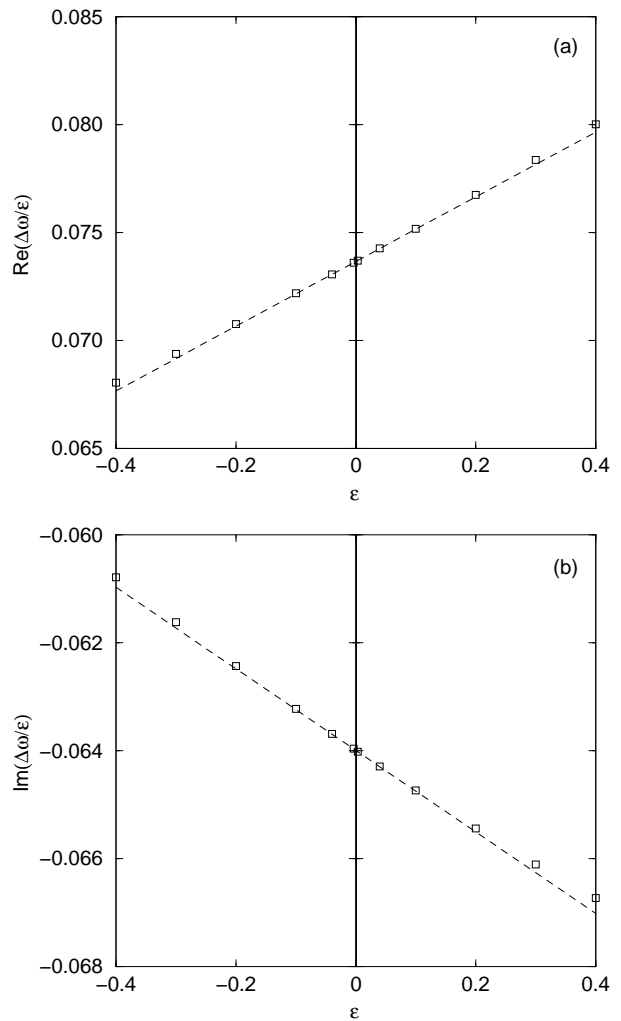}
\caption{The real (a) and imaginary (b) parts of the eigenvalue shift $\Delta\om$ for the system in (\ref{eq:exb01}) (with $\gamma=0.5$), due to a perturbation adding $\epsilon$ to~$K(1,1)$. Straight lines are second-order RSPT; points are roots of the characteristic
polynomial.}\label{fig1}
\end{figure}

Consider the example in Section~\ref{subsect:exb} with $\gamma= 0.5$ for~$\H_0$, and set $K(1,1) \mapsto K(1,1) + \epsilon$. Figure~\ref{fig1} shows $\Delta \om_1(\epsilon)$ for the least damped mode $\om_1 = 2.267 - 0.699 i$. The points are roots of $J(\om)$; the line is given by (\ref{eq:ndp04}), the intercept (slope) verifying its first (second) term. The remaining error (a quadratic in the plot) reveals an $\epsilon^3$ term in~$\Delta\om$. RSPT now gives $\im\Delta\om$ correctly as well.

Re-write the first-order shift allowing arbitrary normalization of $\f_j$, as $\Delta (\om_j^2)/\epsilon =  2\om_j(\Delta K)_{jj}/(\f_j,\f_j )$. Agreement with the conservative limit is now apparent. This also shows that the bilinear map is no mathematical artifact, since it relates directly to the frequency shift---$(\Delta K)_{jj}$ being the expected expression. Incidentally, there could be large shifts if $(\f_j,\f_j )$ is small; the interesting consequences in optics~\cite{excess} are discussed in detail elsewhere~\cite{Pet,pap2}.

Since $\im\Delta \om_j=\sum_n\delta^{(n)}$ ($\delta^{(n)}\propto\epsilon^n$) is novel in the generalized RSPT, one further aspect deserves mention. Suppose the state $j$ is weakly damped with quality factor~$\mathcal{Q}$, but other states are strongly damped: $\gamma_j/\om \sim {\cal Q}^{-1} \ll 1$ and $\gamma'\!/\om \sim 1$; here $\gamma_j \equiv \left|\im\om_j\right|$, $\gamma'$ is the typical $\left|\im\om_k\right|$ for \emph{other}~$k$, and $\om$ is a typical $\left|\re\om_l\right|$. One then has $\delta^{(1)} = \im[\epsilon(\Delta K)_{jj}]\sim\epsilon\gamma_j$ [because $f_j\in\bm{R}$ if $\gamma_j\rightarrow 0$], but $\delta^{(2)}\sim\im[ \epsilon^2(\Delta K)_{kj}^2/(\om_k{-}\om_j)]\sim \epsilon^2\gamma'$. The ratio $\delta^{(2)}\!/\delta^{(1)}$ between energy loss through transitions to other states and direct change to the damping thus is $\sim\epsilon\mathcal{Q}$, which could be \emph{larger} than one if $\mathcal{Q}^{-1}\ll\epsilon\ll1$ (very narrow resonance).

\subsection{Perturbation around level crossings}
\label{subsect:ptd}

Suppose, as parameters are tuned, $M$ eigen\emph{values} merge. Generically, the corresponding eigen\emph{vectors} also merge (criticality), so the remaining ones are incomplete~\cite{pap2,unusual}. Instead, here we consider level crossing, i.e., at an $M$th-order zero of $J(\om)$ there are $M$ independent eigenvectors. Again exactly as in the conservative case, the (typically small) $M \times M$ block is explicitly diagonalized, leaving a perturbation without intra-block elements. Thus zero denominators $\om_k - \om_j$ are avoided, and the formulas in Section~\ref{subsect:ptnd} apply; we omit the details. The same method avoids small denominators \emph{near} a level crossing, where a group of eigenvalues (but not the eigenvectors) are close.

Near-crossing leads to level repulsion and attraction~\cite{repul}, which can now be stated precisely. Focus on near-degenerate states $j=1,2$ and remove the average eigenvalue. In the associated subspace, $(\bar{\H}_0)^\bdot{}_\bdot = \mathop\mathrm{diag}(\delta,-\delta)$. Perturb this as in (\ref{eq:ndp01}), with
\beq
  \Delta \bar{\H}^\bdot{}_\bdot = \begin{pmatrix} 0 & k \\ k & 0 \end{pmatrix}\;,
\eeql{rep02}
where $k=(\Delta K)_{12}$ as in (\ref{eq:ndp05}). Then the eigenvalues are
\beq
  \om = \pm \sqrt { \delta^2 + (\epsilon k)^2 }
\eeql{eq:rep03}
($|\delta|,|\epsilon|\ll1$, but $\epsilon/\delta$ is arbitrary). Note that the physically relevant quantity is $k^2$, meshing with the normalization (\ref{eq:orth1}), which fixes $k$ up to a sign.

\emph{If} the two modes correspond to two near-identical oscillators (as is always the case for $N=2$), then $f_{1,2}^2$ and hence $k$ are real; cf.\ Section~\ref{subsect:exa}. One gets the oft-quoted~\cite{repul} result that frequencies repel but widths attract. This also holds for two modes which are both near crossing and weakly damped, by (\ref{eq:orth1}) and (\ref{eq:ndp05})~\cite{weak-d}.

However, the general case is not as simple, because $K$ and $\Gamma$ need not commute. For instance, take
\beq
  K=\begin{pmatrix} 2{+}4\delta & 0 & 0 \\ 0 & 2 & 2 \\ 0 & 2 & 6 \end{pmatrix}\;,\qquad
  \Gamma=\begin{pmatrix} 2 & 0 & 0 \\ 0 & 1 & 0 \\ 0 & 0 & 4 \end{pmatrix}\;;
\eeql{tricky}
$\delta$ lifts the degeneracy at $\om=1{-}i$. Perturbation by
\beq
  \Delta K=\begin{pmatrix} 0 & \mu_{12} & \mu_{13} \\ \mu_{12} & 0 & 0 \\
                           \mu_{13} & 0 & 0\end{pmatrix}
\eeq
[with $\mu_{ij}$ real and $O(1)$] yields $\Delta \bar{\H}$ as in (\ref{rep02}), where $k=[\mu_{12}/2-\mu_{13}(1{+}i)/4]\sqrt{2}$ can have \emph{any} phase. Sufficiently near the level crossing---the region of interest---$k$~can be taken constant as $\epsilon$ runs through the reals. Simple geometry then shows that the locus of $\omega$ in (\ref{eq:rep03}) is part (by the restriction $\epsilon^2>0$) of an orthogonal hyperbola, with (un)stable direction $ik$~($k$).


\section{Continuum models}
\label{sect:cont}

The continuum limit $N \rightarrow \nobreak\infty$ is relevant to classical and also quantum (Section~\ref{sect:disc}) \emph{fields}, e.g., electrodynamics in an absorbing medium. Although the examples below are 1-d, the theory also applies to three dimensions.

\subsection{String subject to viscosity}
\label{sect:visc}

Consider the example in Section~\ref{subsect:exc}. In the obvious continuum limit $\Delta \rightarrow 0$, $N \rightarrow \infty$ with $a = (N{+}1)\Delta$ fixed, and without loss of generality scaling $k\Delta^2 = 1$, the dynamics obeys (\ref{eq:stringa02}) for $0 \le x \le a$, with $\phi(0) = \phi(a) = \nobreak0$. This models a string subject to viscosity. (Generalization to $0 \le x < \infty$ or $-\infty < x < \infty$, in both cases with $\phi$ vanishing at both ends, is straightforward.)

In terms of the momentum $\hat{\phi}(x) \equiv \partial_t \phi(x)$ and the two-component field $\bm{\phi} = \bm{(}\phi,\hat{\phi}\bm{)}^\mathrm{T}$, one has $i\partial_t \bm{\phi}= \H \bm{\phi}$, with
\beq
  \H(x) = i \begin{pmatrix} 0 & 1 \\ \partial_x^2 - V(x) & -\Gamma(x) \end{pmatrix}\,.
\eeql{eq:defha}
The bilinear map is (up to an irrelevant overall factor $\Delta$)
\beq
  ( \bm{\psi}, \bm{\phi} ) = i \! \int_0^a \!(\psi\hat{\phi}
  + \hat{\psi}\phi + \psi\Gamma\phi)\, dx\; .
\eeql{eq:wavea04}
Again, $\H$ is symmetric under~(\ref{eq:wavea04}).

The eigenvalue equation is $\H \f_j = \om_j \f_j$, or
\beq
  \left[ -\om_j^2  -i \om_j \Gamma(x)- \partial_x^2 + V(x) \right] f_j(x) = 0\;.
\eeql{eq:wave05}
This \emph{dissipative Sturm--Liouville} (DSL) problem could be generalized further by $-\partial_x^2 \mapsto -\partial_x p(x) \partial_x$ for suitable~$p$. In the familiar case $\Gamma = 0$, each $V$ gives a complete set ${\{ f_j \}}_V$, orthogonal under the conventional inner product. Now, each $(V,\Gamma)$ gives a set ${\{ \f_j \}}_{V,\Gamma}$, orthogonal under (\ref{eq:wavea04}) and expected to be complete, with the convergence of infinite two-component sums guaranteed if $(V,\Gamma)$ are suitably smooth and bounded. Such DSL systems are interesting both for, e.g., functional-analytic aspects such as completeness and convergence, and for the specific orthogonal functions that emerge.

If $\Gamma(x) = 2\gamma$, (\ref{eq:wave05}) becomes $(V-\partial_x^2)f_j = \Omega_j^2f_j$, where $\Omega_j$ and $\om_j$ are related by (\ref{eq:stringa05}), and $V-\partial_x^2$ has a complete real eigensystem $\{\Omega_j^2,f_j\}$. The norm $(\f_j,\f_j)$ equals the conventional one $\int_0^a\! f_j^2 \, dx$ up to a pre-factor $\pm2\sqrt{\smash{\Omega_j^2}{-}\smash{\gamma}^2}$ [cf.~(\ref{eq:stringa06})], which vanishes at critical damping.

All these parallel the discrete version in Section~\ref{subsect:exc}.

\subsection{Waves in open systems}
\label{sect:open}

Next consider a similar string on $0 \le x < \infty$, segmented into the ``system" $x \le a$ and a trivial [$V=\Gamma=0$] ``bath" $x>a$, into which the waves are outgoing:
\beq
  (\partial_t + \partial_x)\phi(x{>}a) = 0\;,
\eeql{OWC}
which follows automatically if the bath is at rest initially. So the ``system" is to be solved with the same outgoing (rather than nodal) condition at $x=a$. For simplicity, also take $\Gamma(x{\le}a)=0$ so that damping occurs only through the escape of waves, not through the field equation. This models a laser cavity with output coupling.

To see how also this model is a limit of (\ref{eq:eqmot1}), set $x=\Delta\alpha$, with $\Delta=a/N$, $0\le\alpha\le N$, and $\phi(0)=0$. For $\alpha<\nobreak N$, the discrete dynamics can be written down immediately. For the last particle, reference to the next one just outside the system can be eliminated by (\ref{OWC}) as $[\phi(N{+}1)-\phi(N)]/\Delta=-d_t\phi(N)$, resulting in
\beq
  \left[d_t^2+\frac{d_t}{\Delta}+V(N)\right]\phi(N)
  =\frac{\phi(N{-}1)-\phi(N)}{\Delta^2}\;.
\eeq
These exactly correspond to (\ref{eq:stringb00})--(\ref{enddamp}) for $k=\Delta^{-2}$ and $\gamma=\Delta^{-1}$---without incoming waves, the outside merely acts as a source of ohmic damping on the right endpoint.

The continuum limit of the bilinear map reads
\beq
  ( \bm{\psi}, \bm{\phi} )
  = i \biggl[\int_0^a \!(\psi\hat{\phi}
  + \hat{\psi}\phi)\, dx + \psi(a) \phi(a) \biggr]
\eeql{eq:waveb04}
(again up to a factor $\Delta$), i.e., (\ref{eq:wavea04}) with $\Gamma(x) = \delta(x{-}a)$. The vital symmetry of $\H$ under (\ref{eq:waveb04}) follows as a limit of (\ref{eq:symh}). For a direct proof integrate the $-\partial_x^2$ term by parts; the resulting surface term cancels the one in (\ref{eq:waveb04}) \cite{openwavermp,twocomp}. All results then carry through as before.

Our extensive study of such open wave systems~\cite{openwavermp} motivates much of the present work (notably the \emph{ansatz} for the bilinear map). One can thus apply eigenexpansions etc.\ to, say, optical cavities~\cite{droplet}. In particular, the diagonal bilinear map, known to Zeldovich~\cite{zel} long ago at least for eigenfunctions, has been used for stationary perturbation theory (\cite{openwavermp}~and references therein).

This previous work points to a remaining subtlety. Namely, while completeness is a simple consequence of linear algebra and counting in the discrete case, the modes of the open cavity  are complete only if $V(x)$ or one of its derivatives (depending on the permissible degree of regularization in the field expansion) are discontinuous at $x=a$---demarcating the cavity. One should thus study how the eigenexpansion behaves as $\Delta\rightarrow0$, depending on whether or not this discontinuity condition is satisfied in the limit [one can speculate that some modes of (\ref{eq:stringb00})--(\ref{enddamp}) run off to $-i\infty$ if it is violated]. Ideally, this should yield a completeness proof alternative to the existing one, based on the analyticity of the continuum Green's function for $\im\om<0$ but limited to 1-d space.

\subsection{Wronskian}

Assuming nearest-neighbor couplings has turned these continuum models into second-order differential equations, for which there is an extra ingredient. One can define, for any $\om$, two solutions $f(x,\om)$ and $g(x,\om)$ satisfying the left and the right boundary condition respectively. For, say, a string fixed at both ends, $f(\om,0)=0$, $g(\om,a)$ = 0. The $x$-independent Wronskian is $J(\om) = g(x,\om)\partial_xf(x,\om) - f(x,\om)\partial_xg(x,\om)$. At a zero $\om_j$ of~$J$, $f \propto\nobreak g$ satisfy \emph{both} boundary conditions, hence are eigenfunctions. Thus, the Wronskian plays the role of the characteristic polynomial. Moreover, the (normalization-independent) Green's function can be represented as $\tilde{G}(x{<}y,\om) = f(x,\om) g(y,\om) / J(\om)$ [$\tilde{G}(y,x)=\tilde{G}(x,y)$].

There are however differences from the general finite-dimensional case: 1-d second-order differential equations do not admit level crossing, so a multiple zero of $J$ always implies criticality. Also, the one-sided function $f$ (or $g$) can be studied \emph{near}~$\om_j$, of use for critical points~\cite{JB}.


\section{Discussion}
\label{sect:disc}

In the conservative case, a classical system's eigenvectors (e.g., electromagnetic \emph{plane} waves) are a starting point for discussing nonlinear phenomena (e.g., second-harmonic generation), quantization (e.g., plane-wave creation and annihilation, Feynman propagators), as well as quantum interactions (e.g., a vertex in a Feynman diagram linking several propagators). Similarly, our eigenexpansion for ohmically damped \emph{linear classical} oscillators is relevant for nonlinear and/or quantum systems as well---especially useful for systems with many degrees of freedom dominated by a few modes (typically at low temperatures or when some masses are large). Here we only provide a sketch; details will be given elsewhere.

\subsection{Nonlinear oscillators}
\label{subsect:nonlin}

Consider for example a set of oscillators described by
\beq
  d_t^2 \phi(\alpha) + [\Gamma(\alpha,\beta) d_t
  {+} K(\alpha,\beta) 
  {+}\lambda(\alpha,\beta,\gamma)\phi(\gamma)] \phi(\beta) = 0\;.
\eeq
Using the expansion $\bm{\phi} (t) = \sum_j a^j(t) \f_j$, one shows that
\begin{gather}
  (d_t + i\om_j ) a^j(t) = -i \lambda_{jkl} \, a^k(t) a^l(t)\;,\label{eq:nonlin03}\\
  \lambda_{jkl} = \lambda(\alpha,\beta,\gamma) 
  f_j(\alpha) f_k(\beta) f_l(\gamma)\;.\label{eq:nonlin04}
\end{gather}
For small $\lambda$, these can be solved perturbatively. In particular, (\ref{eq:comp0a}) allows $a^j(0)$ to be found by projecting $\bm{\phi}(0)$.

\subsection{Quantum and thermal correlation functions}
\label{subsect:quant}

To deal with quantum and/or thermal physics (for the moment ignoring nonlinearities), we need to
modify two ingredients.  First, the equation of motion becomes 
\beq
  d_t^2 \phi(\alpha) + \Gamma(\alpha,\beta) d_t \phi(\beta)
  + K(\alpha,\beta) \phi(\beta) = \eta(\alpha,t)\;.
\eeql{eq:quan01}
The noise $\eta$ coming from the bath satisfies the fluctuation--dissipation theorem~\cite{fd}
\beq
  \langle\tilde{\eta}(\alpha,\om)\eta(\beta)\rangle=
  \frac{2\om}{1-e^{-\om/T}}\Gamma(\alpha,\beta)\;,
\eeql{F-D}
where $\langle \cdots \rangle$ denotes the expectation value at temperature~$T$, and $\hbar=k_{\rm B}=1$. For finite-$T$ \emph{classical} dynamics, the rhs reduces to $2T\Gamma(\alpha,\beta)$. In two-component form,
\beq
  (i d_t - \H ) \bm{\phi} = i \bm{S}(t)\;,\qquad
  \bm{S} = \begin{pmatrix}0 \\ \eta\end{pmatrix}\;.
\eeql{eq:quan03}
In the eigenvector basis~\cite{Ho},
\beq
  (d_t + i \om_j ) a^j = S^j(t)\equiv \bm{(}\f_j , \bm{S}(t)\!\bm{)}
  = \langle \f^j | \bm{S}(t) \rangle\;.
\eeql{eq:quan05}
Correlators such as $\langle a^j(t)a^k\rangle$ are then related to $\langle S^j(t)S^k\rangle$, which can be obtained from (\ref{F-D}) and (\ref{eq:quan03}).

Classically, (\ref{eq:quan01})--(\ref{eq:quan05}) can already yield nontrivial Langevin dynamics. In quantum mechanics, $\bm{\phi}$ and hence its coefficients $a^j$ become operators.  Commutators are evaluated by first substituting $a^j = ( \f_j, \bm{\phi} )$ [implying $a^{-j}=\mp i(a^j)^{\dagger}$ when combined with (\ref{eq:norm3})], and then imposing the canonical $[ \phi(\alpha) , \hat{\phi}(\beta) ] = i \delta(\alpha,\beta)$. One finds
\beq
  [a^j,(a^k)^{\dagger}]
  =(\om_j + \om_k^*) f_j(\alpha) f_k^*(\alpha)\equiv\Delta^{jk}\;.
\eeql{eq:quan13}
For $\Gamma \rightarrow 0$, the orthogonality under (\ref{eq:blmap}) is $(\om_j{+}\om_k)\* f_j(\alpha) f_k(\alpha) = \delta^{jk}$, with all quantities real. Hence, $\Delta^{jk} = \delta^{jk} + O(\Gamma)$ so that $a^j$ and $(a^j)^{\dagger}$ generalize the usual annihilation and creation operators. Crucially, however, their commutators (\ref{eq:quan13}) are not diagonal---rooted in the very nature of the bilinear map, and implying that the Feynman propagator is likewise non-diagonal. Diagrammatic expansions will thus show interesting features.

In fact, we can calculate the correlation function
\beq
  F(\alpha,\beta,t) = \langle \phi(\alpha,t) \phi(\beta,0) \rangle
\eeql{eq:quan31}
in a way that bypasses some of these technical complications. For $\om\in\bf R$, from general considerations~\cite{fd}
\beq
  {\tilde F}(\alpha,\beta,\om) =
  2{(1-e^{-\om/T})}^{-1} \im \tilde{G}^{QP} (\alpha,\beta,\om)\;.
\eeql{eq:quan32}
Using (\ref{eq:green03}) for $\tilde{\cal G}$ then gives
\beq
  {\tilde F}(\alpha,\beta,\om) =
  \frac{2i}{1-e^{-\om/T}} \sum_j \frac{\om f_j(\alpha)f_j(\beta)}{\om^2-\om_j^2}\;,
\eeql{eq:quan35}
with poles at both the $\om_j$ and the Matsubara frequencies $\om_m = 2m\pi i T$. Substituting the integrated (\ref{eq:quan05}) into the eigenexpansion of (\ref{eq:quan31}) also yields (\ref{eq:quan35}). Again, we omit details, and only stress that the eigenbasis provides a direct correspondence with the frequency singularities.

The Feynman propagator can be evaluated in terms of $F$ and~$\mathcal{G}$. The perturbative treatment of interactions of Section~\ref{subsect:nonlin} (readily promoted to the operator level) then opens the way to a diagrammatic analysis.

\subsection{Conclusion}
\label{subsect:concl}

We have established an eigenvector expansion for a system of ohmically damped oscillators such that almost all the usual results for conservative systems carry over. In a sense, introducing the bilinear map is equivalent to inverting the (potentially huge) matrix $\langle \f_j | \f_k \rangle$. Initial-value problems are readily solved, and RSPT is valid for the \emph{complex} eigenvalues. The possible extension to nonlinear and quantum phenomena has been sketched. The present formalism fails when any mode is critically damped; this is handled in the sequel~\cite{pap2}.

These developments should be useful whenever there is (a)~dissipation and (b)~many degrees of freedom.  Examples would include mechanical vibrations, acoustics, electromagnetism in absorbing (as well as gain) media, large Josephson-junction arrays and macromolecule vibrations in a viscous medium. 

An extension to non-ohmic damping (which, by the Kramers--Kronig relations, typically would entail dispersion in the force constants as well) would enable modeling the many dispersive systems encountered in applications. Moreover, a closer study of the quantum dynamics requires a high-frequency cutoff on the damping.


\acknowledgments

This work is built upon a long collaboration with E.S.C. Ching, H.M. Lai, P.T. Leung, S.Y. Liu, W.M. Suen, C.P. Sun, S.S. Tong, and many others.  KY thanks R.K. Chang for discussions on microdroplet optics, initiating our interest in waves in open systems, and K.T. Chan, M.~Sainsbury, and N.~Stephen for discussions on mechanical vibrations. We thank S.L. Cheung for discussions and help with the figure. AMB was supported by a C.N. Yang Fellowship.


\appendix

\section{Fast modes}
\label{sect:constr}

\subsection{Equations of motion}
\label{subsect:constreq}

If one particle is much heavier than all others, the latter can be treated as ``fast modes", the momentum of which is eliminated. In preparation for such a study and for another application in Appendix~\ref{subsect:constrex}, here we consider \emph{one} fast mode $\alpha=N{+}1$ of mass $\epsilon\ll1$. Specifically,
\beq
  d_t^2\phi(\alpha)+[\Gamma(\alpha,\beta)d_t{+}K(\alpha,\beta)]\phi(\beta)=-A(\alpha)\phi(N{+}1)
\eeql{fast1}
for $1\le\alpha\le N$ (also the range of the summation convention), while the light particle is coupled to the others only by force constants $A(\alpha)$, not by cross-damping~\cite{cross-d}:
\beq
  (\epsilon d_t^2+\gamma d_t+\kappa)\phi(N{+}1)=-A(\alpha)\phi(\alpha)\;.
\eeql{fast2}
This model (easily extended to more light masses) is contained in (\ref{eq:eqmot1}), if one rescales $\underline{\phi}(N{+}1)\equiv\sqrt{\epsilon}\phi(N{+}1)$.

In the adiabatic limit $\epsilon\rightarrow0$, the last particle's momentum follows the other variables:
\beq
  d_t\phi(N{+}1)=-c\phi(N{+}1)-B(\alpha)\phi(\alpha)\;,
\eeql{fast3}
with $c=\kappa/\gamma$ and
\beq
  B(\alpha)=\gamma^{-1}A(\alpha)\;.
\eeql{fast4}
The dynamical system defined by (\ref{fast1}), (\ref{fast3}), and~(\ref{fast4}) is dissipative even if $\Gamma=0$: (\ref{fast3}) breaks invariance under $t\mapsto-t$. The system is stable if the original one is, i.e., if
\beq
  {\cal K}\equiv\begin{pmatrix} K & A \\ A^{\rm T} & c\gamma\end{pmatrix}>0\;.
\eeql{fast5}
Assuming $K>0$, Sylvester's theorem~\cite{biorth} guarantees stability  if $\det{\cal K}>0$; in particular, one needs $c\gamma>0$. Since $\gamma$ is taken positive in (\ref{fast2}) one finds that $c>0$, which is plausible given the form of (\ref{fast3}).

On the Hamiltonian level, (\ref{fast3}) means that $\hat{\phi}(N{+}1)$ is not dynamical but obeys a constraint instead, hence the phase space has $2N{+}1$ dimensions. Thus, define
\beq
  \bm{\phi} =
  \bm{(} \phi(1) , \ldots , \phi(N) ;
  {\hat \phi}(1) , \ldots , {\hat \phi}(N) ; \phi(N{+}1) \bm{)}^{\rm T}\;.
\eeql{eq:constr04}
Then the dynamics takes the standard form (\ref{eq:eqmot3}), with
\beq
  \H^{\bdot}{}_{\bdot} = i \begin{pmatrix}
  0 & I & 0 \\ -K & -\Gamma & -A \\ -B^{\rm T} & 0 & -c\end{pmatrix}\;.
\eeql{eq:constr05}


\subsection{Bilinear map}
\label{subsect:constrblm}

Since the above is merely a limiting case of our general model, the bilinear map should be
\pagebreak
\begin{align}
  (\bm{\phi},\bm{\psi})&=i\bigl[\phi(\alpha)\hat{\psi}(\alpha)+\hat{\phi}(\alpha)\psi(\alpha)
    \notag\\ &\phantom{=i\{}+\underline{\phi}(N{+}1)\hat{\underline{\psi}}(N{+}1)
    +\hat{\underline{\phi}}(N{+}1)\underline{\psi}(N{+}1)\notag\\
    &\phantom{=i\{}+\phi(\alpha)\Gamma(\alpha,\beta)\psi(\beta)
    +\underline{\phi}(N{+}1)\epsilon^{-1}\gamma\,\underline{\psi}(N{+}1)\bigr]\notag\\[.5mm]
  &=i[\phi(\alpha)\hat{\psi}(\alpha)+\hat{\phi}(\alpha)\psi(\alpha)
        +\phi(\alpha)\Gamma(\alpha,\beta)\psi(\beta)\notag\\
    &\phantom{=i\{}+\phi(N{+}1)\gamma\psi(N{+}1)]+{\cal O}(\epsilon)\label{fast6}
\end{align}
in terms of the vectors (\ref{eq:constr04}). By (\ref{eq:defg3}), this leads to
\beq
  g_{\bdot\bdot} = i\begin{pmatrix} \Gamma & I & 0 \\ I & 0 & 0 \\
     0 & 0 & \gamma\end{pmatrix}\,,\quad
  \H_{\bdot\bdot} = \begin{pmatrix} K & 0 & A \\ 0 & -I & 0 \\
     A^{\rm T} & 0 & c\gamma \end{pmatrix}\,.
\eeql{eq:constr10a}
Interestingly, when starting out by \emph{postulating} (\ref{fast3}) in addition to (\ref{fast1}), the proportionality (\ref{fast4}) follows from demanding that (\ref{eq:constr10a}) be symmetric. Likewise, requiring the existence of a non-increasing energy function bounded from below then leads to~(\ref{fast5}).

The adiabatic limit can similarly be taken in essentially all results of the previous sections, yielding a realization of our framework in an odd-dimensional phase space.


\subsection{Example}
\label{subsect:constrex}

Because of (\ref{fast4}), we should demonstrate nontrivial examples.  Consider a string of particles $0\le\alpha\le N{+}1$, spaced by $\Delta$, with $a = (N{+}1)\Delta$ and $\phi(0) = 0$. Let the force constants be given by (\ref{eq:stringa01}) with $k\Delta^2=1$, while for simplicity $\Gamma = 0$.  The nearest-neighbor couplings (\ref{eq:stringa01}) also act between particles 0 and~1, and between particles $N$ and $N{+}1$.  The latter implies $A(\alpha) = -\Delta^{-2} \delta(\alpha,N)$.

The right boundary condition is the discrete version of (\ref{OWC}), namely $d_t \phi(N{+}1) = -\Delta^{-1} [ \phi(N{+}1) - \phi(N)]$. Compared to (\ref{fast3}), we have $B(\alpha)=-\Delta^{-1}\delta(\alpha,N)$ and $c=\Delta^{-1}$. Crucially, (\ref{fast4}) is satisfied, with $\gamma = \Delta^{-1}$.

At least formally, for $N\rightarrow\infty$ (with $a$ fixed) one retrieves the model of Section~\ref{sect:open}. Indeed, the continuum limit of (\ref{fast6}) agrees (up to an overall factor of~$\Delta$) with (\ref{eq:waveb04}). It would be instructive to compare the convergence of the present discrete model with the more physically motivated one from the main text.



\begin{thebibliography}{99}
\bibitem{mass}
Any symmetric mass matrix $\mathcal{M}$ can be diagonalized and scaled, so the $d_t^2$ term is in fact general. Under a further orthogonal transformation (not affecting $\mathcal{M}=I$), either $K$ or $\Gamma$ (but in general not both) can be diagonalized.
\bibitem{damp}
This is readily demonstrated as follows. Multiply the defining equation by $f_j(\alpha)^*$ and sum over $\alpha$. Define $q = f_j(\alpha)^* f_j(\alpha) > 0$, $\gamma = f_j(\alpha)^* \Gamma(\alpha,\beta) f_j(\beta) / 2q \ge 0$ and $k = f_j(\alpha)^* K (\alpha,\beta) f_j(\beta) / q > 0$, giving $\om_j^2 +2i\om_j \gamma - k = 0$ familiar from the case of a single oscillator.
\bibitem{zero-pair}
Since the total number of modes is even, for these systems in a certain sense even zero-modes are paired.
\bibitem{timescale}
For dimensional reasons, the product of the momenta should carry a factor $\tau^2$, where $\tau$ is a time scale.  For simplicity however we take $\tau=1$ throughout.
\bibitem{real}
Although the physical coordinates and momenta are real, frequency eigenvectors are complex.
\bibitem{deg-phase}
For off-axis level crossings, (\ref{eq:norm3}) can be defined to hold by choosing the two sets of eigenvectors consistently. For simple zero-modes, (\ref{eq:norm3}) is a symmetry; comparison with the overdamped case of (\ref{eq:exa05}) shows that both signs can occur. Finally, for crossing zero-modes, both (\ref{eq:eigenright}) and (\ref{eq:orth}) are real, so that there exists a real (up to a prefactor) basis in which each vector maps to itself under (\ref{eq:norm3}). Stronger statements are not possible. Take, e.g., $K=\diag(3,15)$ and $\Gamma=\diag(4,8)$. At $\om=-3i$, an eigenbasis of the mentioned type reads $\f_1=\frac{1}{2}\sqrt{2}e^{i\pi/4}(1,0,-3,0)^{\rm T}$, $\f_2=\frac{1}{2}\sqrt{2}e^{-i\pi/4}(0,1,0,-3)^{\rm T}$, while (\ref{eq:norm3}) maps the orthonormal basis vectors $\bm{g}_1=\frac{1}{2}e^{i\pi/4}(1,i,-3,-3i)^{\rm T}$, $\bm{g}_2^\phs=\pm i\bm{g}_1^*$ to each other. Conversely, at $\om=-3i$ the system $K=3I$, $\Gamma=4I$ of two identical oscillators has an eigenbasis of the first but not of the second type.
\bibitem{notation}
For simplicity, the representation in which various matrices and vector components are expressed will only be explicitly indicated if there is danger of confusion.
\bibitem{alternate}
Because of the sum rules, it is possible to write RSPT in several forms. E.g., by (\ref{eq:ndp05}), the second-order correction to~$\om_j$ involves $\sum_{k \neq j}(\Delta K)_{jk}(\om_j{-}\om_k)^{-1}(\Delta K)_{kj}=f_j(\alpha)\*\Delta K(\alpha,\beta)\* T_j(\beta,\gamma)\*\Delta K(\gamma,\delta)\*f_j(\delta)$, where $T_j(\beta,\gamma)\equiv\sum_{k \neq j} f_k(\beta)\*(\om_j{-}\om_k)^{-1}f_k(\gamma)=\sum_{k \neq j} f_k(\beta)\*\om_k\*[\om_j(\om_j{-}\om_k)]^{-1}f_k(\gamma)\linebreak-f_j(\beta)f_j(\gamma)/\om_j$,
by (\ref{eq:sr1}). Especially for large~$N$, one of these forms will be more convenient computationally. In the above example, the original form is better in that the summand decreases faster for large~$\om_k$.
\bibitem{unusual}
Merging of $M$ eigenvalues first of all implies that $J(\om)$ has an $M$th order zero. If there are two independent eigenvectors $\bm{v}_1, \bm{v}_2$, then by taking linear combinations one can find a $\bm{v}$ whose (say) first component is zero.  Then $\bm{v}$ is not only a zero eigenvector of $Y = \H - \om$, but also of its $(1,1)$ minor matrix $Z$, so that $\det Z =0$ as well. Thus, level crossing requires extra conditions, and is therefore unusual---readily verified in examples.
\bibitem{weak-d}
The latter case leading to (approximately) real $k$ in fact is related to the former. By~\cite{damp}, the two modes can only be weakly damped if $\Gamma$ is small in the associated subspace. Then, roughly, $[K,\Gamma]$ must be small there as well, so that the subtleties mentioned below (\ref{tricky}) do not occur.
\bibitem{cross-d} A cross-damping would yield an extra term $-B'(\alpha)d_t\phi(\alpha)$ on the right of (\ref{fast3}), which however can be removed by $\phi(N{+}1) + B'(\alpha) \phi(\alpha) \mapsto \phi(N{+}1)$ and $B(\alpha)+cB'(\alpha)\mapsto B(\alpha)$.
\end{thebibliography}
\end{document}